\documentclass[12pt]{article}

\usepackage{authblk}

\title{Peristaltic pumping in short annular geometries: An experimental approach for studying Glymphatic flow}
\date{}
\author[1]{Shahaf Ella Salach}
\author[1]{Ron Shnapp}
\affil[1]{Mechanical Engineering Department, Ben-Gurion University of the Negev, Beer-Sheva, Israel}

\usepackage{amsmath}
\usepackage[normalem]{ulem}
\usepackage{graphicx}
\usepackage{subcaption}
\usepackage{setspace}
\usepackage[a4paper, margin=1in]{geometry}
\usepackage[hidelinks]{hyperref}
\hypersetup{
    colorlinks = true,
    citecolor = {blue},
    linkcolor = {blue},
    urlcolor={blue}
}
\usepackage{xcolor}
\usepackage{bm}
\usepackage{amssymb}
\usepackage{float}

\renewcommand{\vec}[1]{\boldsymbol{#1}}            
\newcommand{\av}[1]{\left\langle{#1}\right\rangle} 

\begin{document}

\renewcommand{\labelitemi}{$\bullet$} 
\renewcommand{\labelitemii}{$\Diamond$} 
\onehalfspacing
\pagenumbering{gobble}

\maketitle

\begin{abstract}
Peristaltic pumping is hypothesized to drive fluid transport in several physiological systems, including cerebrospinal fluid flow through cerebral perivascular spaces (PVSs). Cerebral PVSs are unique in the context of peristaltic pumping because they have annular geometry and are orders of magnitude shorter than the peristaltic wavelength. Due to these features, questions were raised as to whether peristaltic pumping is possible under such conditions, and experimental tests for this concept are lacking. This work presents a novel experimental setup that enables direct, detailed measurements of peristaltic flow in short annular channels formed between a compliant inner tube and a rigid outer tube. A propagating pulse wave along the inner tube generates back and forth fluid motion in the annular gap, which we measure using particle tracking velocimetry in a refractive-index matched setup. Despite the instantaneous back and forth motion, net axial fluid transport in the direction of wave propagation is observed, and the resulting net velocity profiles collapse across a range of wall deformation amplitudes. These results provide experimental evidence for net transport induced by long wave length peristaltic deformations in a physiologically relevant flow regime.
\end{abstract}

\section{Introduction}

Fluid transport is ubiquitous in physiological systems, such as blood flow through the cardiovascular system~\cite{ku_blood_1996, milnor1972pulsatile} or lymph circulating through the lymphatic system~\cite{scallan2016lymphatic}, however the mechanisms driving these fluids are not always clear. One type of such driving mechanisms is called peristaltic pumping---a mechanism in which fluid flow is induced by periodic wall deformations that propagate as travelling waves \cite{pozrikidis_study_1987}. 
Peristaltic pumping drives, for example, flow inside the ureter~\cite{Canda2007}, nutrients along the gastrointestinal system \cite{Sinnott2017, mekheimer_peristaltic_nodate} and Cerebrospinal Fluid (CSF) flow through Perivascular Spaces (PVS) in the central nervous system~\cite{iliff_cerebral_2013, mestre_flow_2018, kelley_cerebrospinal_nodate, bilston_arterial_2003}. In the context of the latter, CSF flows inside cerebral PVSs, which are annular channels surrounding small blood vessels~\cite{jessen_glymphatic_2015, wardlaw_perivascular_2020}; there, cardiac-pulsation-derived deformations of the vessel walls are hypothesized to be the driver of CSF flux~\cite{iliff_cerebral_2013, mestre_flow_2018}. This system, often referred to as the glymphatic system, is thought to provide a pathway for nutrients delivery and metabolic waste evacuation for deep brain tissues \cite{iliff_paravascular_2012}. Despite the critical importance of the glymphatic system for normal brain function, the underlying flow mechanism remains unclear, as reviewed in~\cite{kelley_cerebrospinal_nodate, Bohr2022}.

Peristaltic flow has been extensively investigated, predominantly in circular channels and tubes with varying characteristics. Theoretical analysis of peristaltic flow in an axisymmetric domain revealed that the flow rate is proportional to the wave velocity and its dependence on the wave amplitude varies with the ratio between the wave amplitude to the radius of the circular channel~\cite{barton1968peristaltic, shapiro_peristaltic_1969, pozrikidis_study_1987}. The flow rate increases with wave amplitudes, quadratically for small amplitudes and linearly for larger amplitudes.

\begin{figure}[h]
    \centering
    \includegraphics[width=0.9\textwidth]{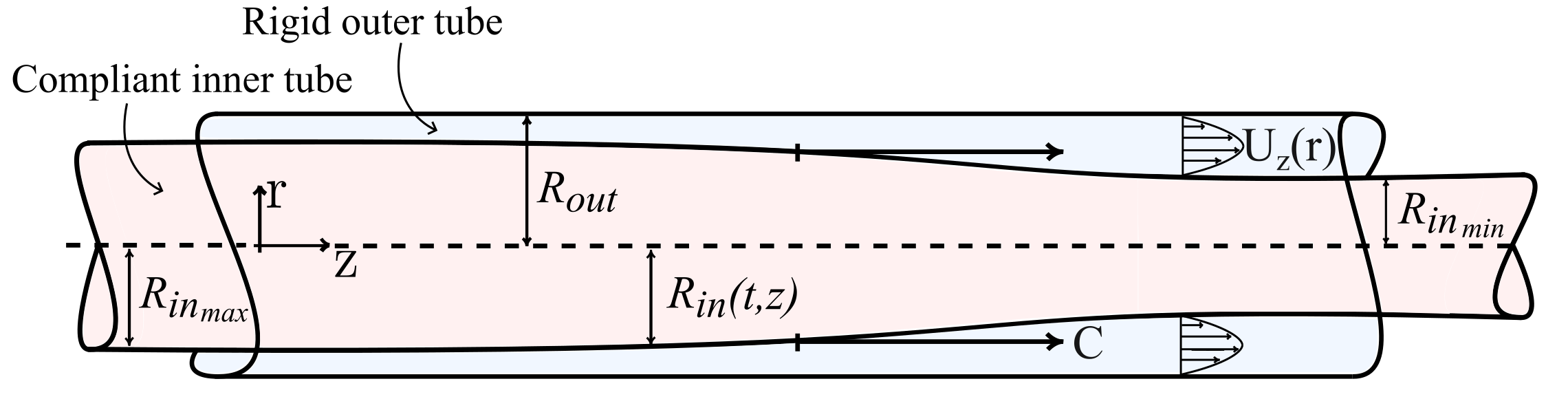}
    \caption{Schematic representation of the flow domain. The annular channel is bounded by an inner compliant tube with a radius of $R_{in}(t,z)$ and an outer rigid tube with a radius of $R_{out}$. $R_{in_{max}}$ and $R_{in_{min}}$ are the maximal and the minimal values of $R_{in}$, respectively. The pulse wave, with a wavelength longer than the flow domain, travels at velocity $C$ and the axial velocity of the fluid in the annular channel is $U_{z}(r)$. The walls of both tubes assumed to be impermeable.}
    \label{fig:problem-scheme}
\end{figure}

In cerebral PVS, the flow is characterized by low Reynolds numbers, annular geometry and very long wavelengths, namely the length of the flow domain is much smaller than the wavelength~\cite{kelley_cerebrospinal_nodate, thomas_fluid_2019}. Furthermore, wall deformations in this system occur on the internal wall of the annular channel (Fig.~\ref{fig:problem-scheme}). Wang and Olbricht~\cite{wang_fluid_2011} developed a theoretical model for fluid flow within porous PVS, assuming sinusoidal peristaltic waves and lubrication approximation. The expression derived for the time-averaged flow rate shows a linear dependence of the flow rate on wave velocity and a quadratic dependence on the outer radius of the annulus. For a case where there is no axial pressure difference and the wave amplitude is small compared to the outer radius, the flow rate increases quadratically with the wave amplitude.

An additional feature of cerebral PVS, evident in in-vivo studies~\cite{iliff_paravascular_2012, vinje_brain_2021}, is that the annular geometry is often not axisymmetric due to the presence of an outer elliptical wall and eccentricity.
This feature was shown to have a profound impact on the peristaltic flow therein. Carr et al.~\cite{carr_peristaltic_2021} showed through numerical simulations that peristaltic flow in asymmetric annular channels is three dimensional with secondary motions in the azimuthal direction. They also found that the resultant time-averaged volumetric flow rate is in the direction of wave propagation. Coenen et al.~\cite{coenen_lubrication_2021} derived a closed-form analytic expression for the mean flow rate in a non-axisymmetric annular tube, which is in a good agreement with the numerical results of Carr et al.~\cite{carr_peristaltic_2021}. This unique geometry was also found to have a lower hydraulic resistance compared to the symmetric case, leading to faster flow of CSF~\cite{tithof_hydraulic_2019}. 

It is important to note that all of the aforementioned theoretical and numerical models adopt a short-wavelength assumption, in which the wavelength of the peristaltic deformation is small compared to the length of the flow domain. While this assumption enables significant simplifications, it does not accurately reflect physiological conditions in cerebral PVS. The wavelength of the cardiac-induced wall deformations is estimated in the range of [0.2, 1] m~\cite{thomas_fluid_2019, romano_peristaltic_2020}, resulting in wavelengths roughly four orders of magnitude larger than the flow domain length~\cite{Bohr2022}. Under these conditions, the blood vessel wall undergoes nearly synchronous expansion and contraction over most of the wave period. Consequently, it is unclear whether peristaltic pumping can serve as the driving mechanism of CSF flow in cerebral PVS.

Our survey of the existing literature on the mechanisms driving fluid flow in cerebral PVSs has also revealed a critical gap: works that sought to elucidate the flow mechanisms are theoretical or numerical, with one experimental work that used a planar (not annular) geometry \cite{coloma_boundary_2019}. We believe that addressing this gap is crucial, as experimental work will allow substantiating the important conclusions found in previous studies, as well as possibly discovering new unexpected phenomena. The issue in this respect is that the complex geometry of cerebral PVS make it difficult to construct experimental synthetic apparatuses that could emulate the driving mechanism while retaining the ability of reliably measuring the flow therein in detail. Our work strives to address this gap by developing a new system allowing to do just that, while emulating the principal driving mechanism of cardiac pulsation.

In this work, we present a novel experimental approach using refractive-index (RI) matched, elastic materials to study peristaltic flow in an annular domain with impermeable walls, as described in Fig.~\ref{fig:problem-scheme}. The experimental apparatus consists of a compliant inner tube housed within a larger, transparent and rigid outer tube. A pulse wave propagating along the inner tube induces periodic wall deformations, generating peristaltic flow inside the annular domain between the tubes. We use high-speed camera and image analysis to evaluate the pulse wave velocity and its amplitude, as well as Particle Tracking Velocimetry (PTV) to measure the instantaneous- and net- flow velocities inside the annular domain in detail. We found that despite the fact that the length of the annular channel is shorter then the pulse wavelength, net flow that depends on the system parameters persists in the direction of wave propagation.

\section{Methodology}

In this section we describe the new experimental apparatus we have developed in order to simulate and measure peristaltic flow in annular geometry, as well as the measurement techniques for the pulsating internal tube and detailed flow measurements through particle tracking.

\subsection{Experimental apparatus}

\begin{figure}[h]
    \centering
    \includegraphics[width=0.9\linewidth]{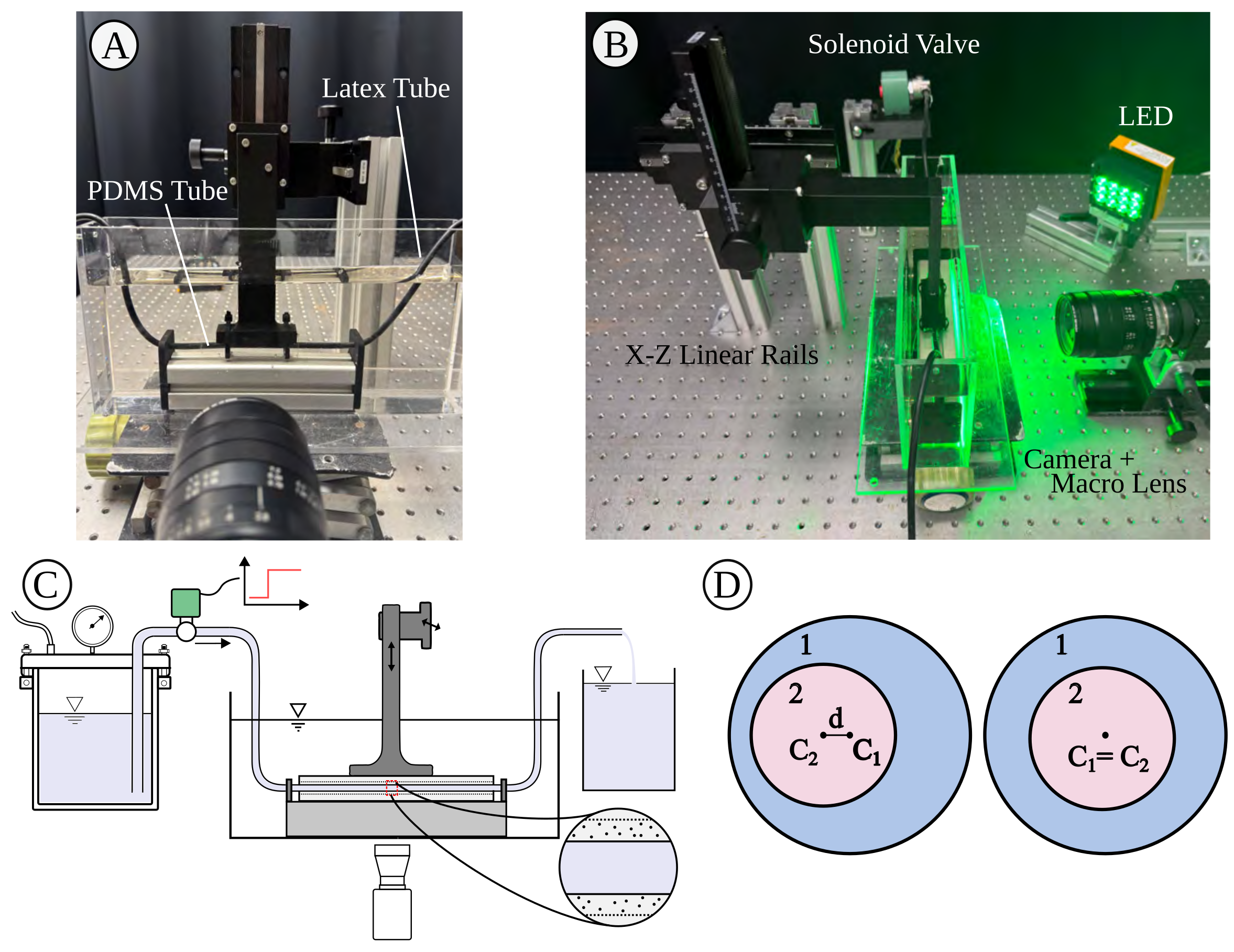}
    \caption{The experimental setup. A) The compliant latex tube is fixed inside the channel and the rigid PDMS tube is held from outside of the channel and fixed to the x-z linear rails. B) The camera is positioned in front of the container, and the compliant tube is connected to a solenoid valve. C) Schematic sketch of the whole experimental setup. D) Cross-section of tubes with eccentricity $d$ (left) and with zero eccentricity where $d=0$ (right). }
    \label{fig:exp-setup}
\end{figure}

The experimental setup consists of an internal compliant tube and an external rigid transparent tube contained in a PMMA open channel, as depicted in Fig.~\ref{fig:exp-setup}A. The height, width, and length of the open channel are $H=150 \, mm$, $W=50 \, mm$, and $L=290 \, mm$, respectively. The internal compliant tube is a 40A-durometer latex rubber tube (McMaster-Carr) with an outer diameter of $6.3 \,mm$ and wall thickness of $0.79 \,mm$. The external rigid tube is made of Polydimethylsiloxane (PDMS) and fabricated in our lab; its inner diameter is $7.8 \, mm$ and its length is $79 \, mm$. The container is filled with a water-glycerin mixture with $60.7w\%$ of glycerin to achieve RI matching with the PDMS. More details about the PDMS fabrication process and RI matching are presented in Section \ref{pdms-RI}.

The internal tube is fixed horizontally inside the channel using a 40$\times$40 mm aluminium profile and 3D-printed fixtures. The external tube is held by a 3D-printed fixture and x-z linear rails that enable us to control the eccentricity between the tubes. A camera (Fastcam Mini AX100, Photron) is positioned in front of the channel and fixed to a linear rail (Fig.~\ref{fig:exp-setup}B), allowing to control the resolution and evaluate the depth of field. A schematic sketch of the complete experimental setup is presented in Fig.~\ref{fig:exp-setup}C. The camera is equipped with a macro lens (Laowa lens 60mm, $\times$2 magnification) for high spatial resolution of approximately $\sim 10 \, \mathrm{\mu m/pixel}$, providing high measurement accuracy. The depth of field was below $1$ mm. The space created between the tubes is annular, and its shape is determined by the eccentricity between the tubes (Fig.~\ref{fig:exp-setup}D).
The experiment described in this paper involved a small degree of eccentricity estimated as $\varepsilon=\frac{d}{R_{in_{0}}} = 0.026 \pm 0.002$, where $R_{in_{0}}$ is the nominal inner radius of the annular channel. We are currently conducting an in-depth study to characterize the effect of the tubes eccentricity.

\subsubsection{Pulse wave generation}

To induce peristaltic flow in the annular gap between the tubes, we applied water pressure pulses through the internal compliant tube, which dilated and contracted in response. The pulses were generated from a pressurized water tank using a solenoid valve (Asco Red Hat 8262H015) controlled by an Arduino Uno R3 microcontroller board. The pulses frequency was set to 1 Hz. The pressure tank was connected to a pressurized air line, and the pressure inside the tank was regulated by a compressed air regulator (McMaster-Carr 8812K52) and measured using a pressure gauge (Cecomp DPG2000). The pressure range of the pulses was $50-190$ kPa. One end of the compliant tube was connected to the solenoid valve and the other end was open to ambient conditions. With a durometer of 40A, which corresponds to Young's modulus of $1.7$ MPa \cite{gent_relation_1958}, the compliant latex tube is highly flexible, expanding its diameter by up to $\sim5 \%$ within the pressure range used in our system.

\subsubsection{PDMS tube fabrication and refractive-index matching} \label{pdms-RI}

PDMS is an elastomeric polymer used in a wide range of applications due to its distinct properties, including optical transparency, elasticity and simple fabrication process \cite{miranda_properties_2022}. The process of PDMS fabrication involves a straightforward procedure in which a PDMS monomer is blended with a curing agent at a typical ratio of 10:1, followed by degassing to eliminate bubbles and subsequently curing, either at room temperature or using an oven. The specific details of this fabrication method affect the PDMS's mechanical and optical characteristics. Generally, for bulk PDMS, the Young's modulus can range between $360$ to $870$ kPa approximately, the Poisson's ratio is $0.5$, and the refractive index is approximately $1.41$ \cite{miranda_properties_2022}.

\begin{figure}[h]
    \centering
    \includegraphics[width=0.7\linewidth]{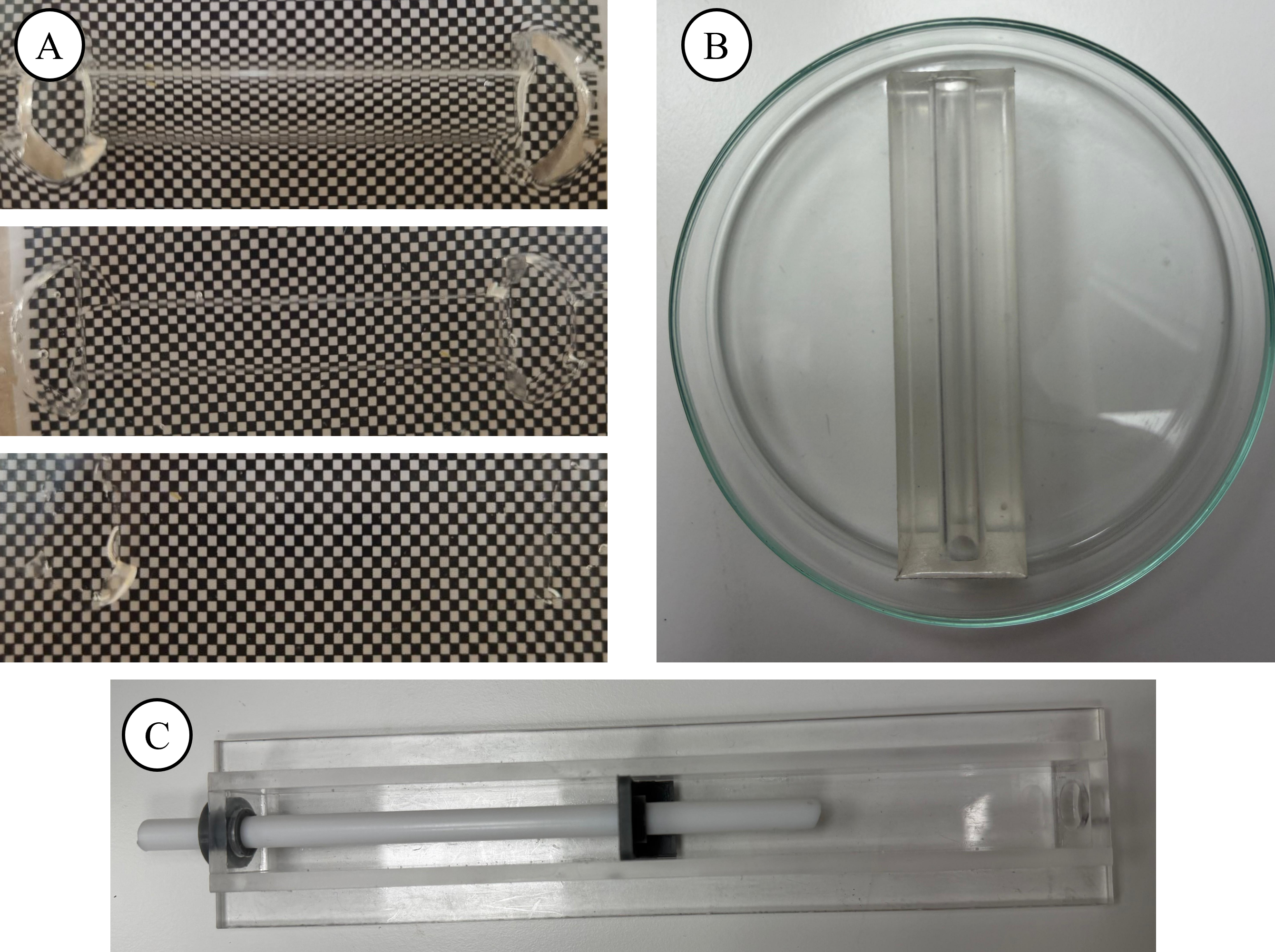}
    \caption{Rigid PDMS tube fabrication and RI matching. A) An optical test of the RI matching between the PDMS and water-glycerin solution used. A cylindrical lens with radius of curvature of 8 mm made of PDMS is immersed in different fluids and placed over a square printed pattern. Top: air; middle: water; bottom: water-glycerin solution at $60.7w\%$ concentration. B) The rigid PDMS tube we fabricated in our lab. C) The Perspex and PTFE stencil used to fabricate the rigid PDMS tube.}
    \label{fig:pdms_tube}
\end{figure}

For our purpose, we used Sylgard® 184 PDMS to create a transparent tube with a size and shape of our choice. Since our focus is solely on the flow between the tubes, the PDMS should have a circular internal geometry, whereas its exterior can be square. Therefore, we designed a PMMA stencil shaped as a rectangular prism with one hole in each end (Fig.~\ref{fig:pdms_tube}). The internal tube wall was formed using a $8$ mm diameter polytetrafluoroethylene (PTFE) rod, due to its non-stick properties \cite{ashokkumar_evaluating_2011}. After mixing the two parts of the PDMS and degassing, the mixture was poured to the PMMA stencil and then cured in a drying oven for 4 hours in $60^\circ$C. The PDMS was then extracted from the template and the PTFE rod was removed. The resulting PDMS tube is presented in Fig.~\ref{fig:pdms_tube}. The shrinkage rate of PDMS during curing is in the range of $0.5-2.5\%$, depending on curing temperature, PDMS components ratio and layer thickness \cite{lee_shrinkage_2008}. Therefore, the actual inner diameter of the PDMS tube is slightly different from the diameter of the PTFE rod, and it was evaluated based on the PTV images.

Optical transparency of the external PDMS tube is essential for reliable results, and can be achieved with RI matching between the tube and the measurement fluid. We measured the RI of our fabricated PDMS tube by measuring the diffraction of a laser beam which was found to be $1.4052\pm0.0545$. Burgmann et al.~\cite{burgmann_refractive_2009} report that a water-glycerin mixture with $60.7w\%$ can meet the requirement for RI matching with PDMS, and we confirmed it with a PDMS concave lens we fabricated at our lab (Fig.~\ref{fig:pdms_tube}A). The RI of the mixture was measured with a refractometer (Abbemat3001, Anton Paar) and found to be $1.4154\pm 0.0001$ at room temperature.

\subsection{Measurements and data evaluation}

\subsubsection{Flow velocity}

\begin{figure}[h]
    \centering
    \includegraphics[width=0.4\linewidth]{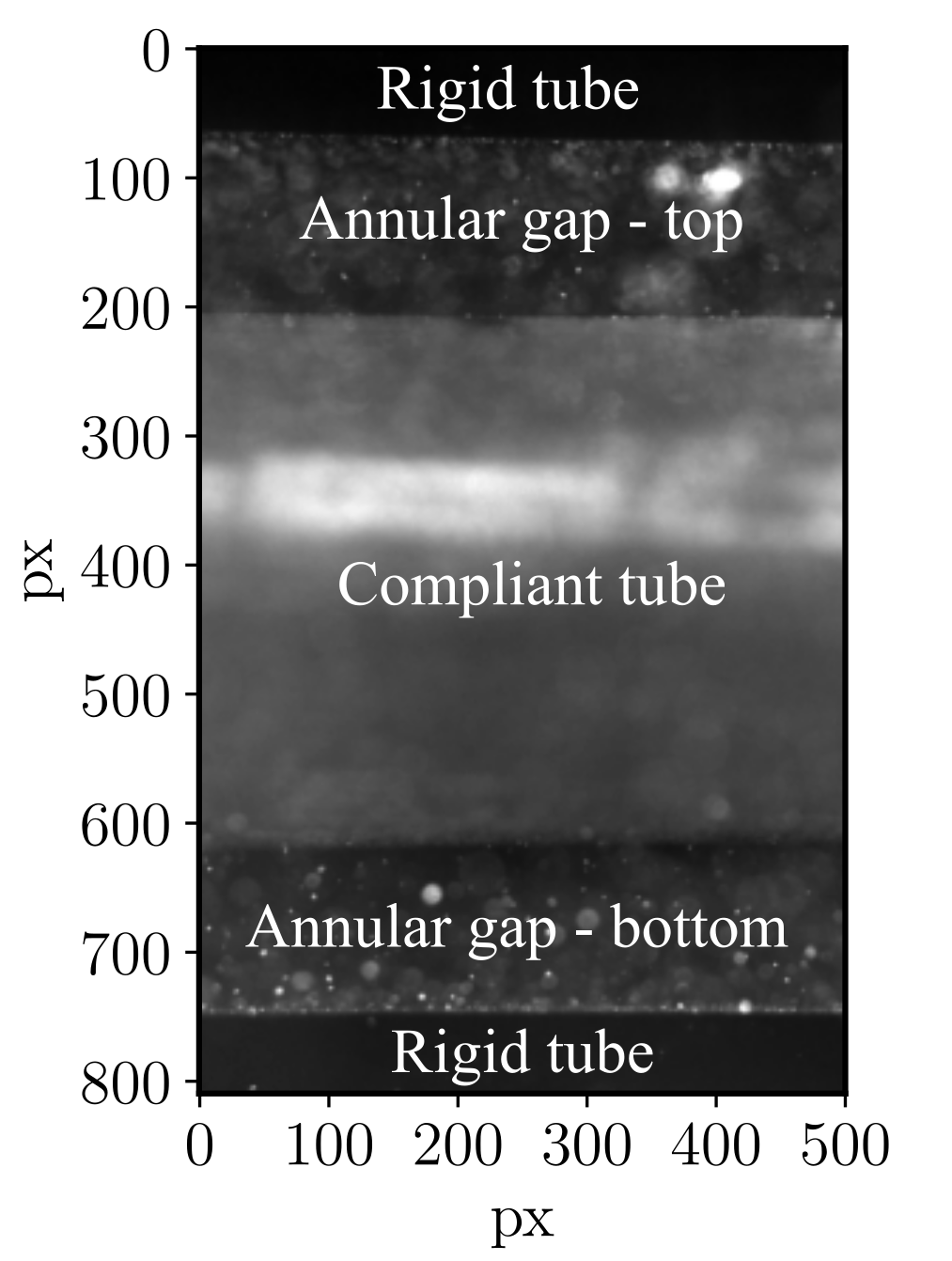}
    \caption{Raw image from the experimental apparatus, showing the inner compliant tube, the annular gap and the external rigid tube.}
    \label{fig:raw_tubes_image}
\end{figure}

The flow velocity was measured with a PTV method using a high-speed camera and LED illumination. In order to accurately capture the entire dilation-contraction cycle, the camera frame rate was set to $250$ frames per second. The flow was evaluated by tracking silver coated Hollow Glass Spheres (S-HGS) with a mean diameter of $10\,\mu$m and density of $1.4 \, \mathrm{g/cm}^3$ (Dantec Dynamics), which were added to the water-glycerin mixture prior to the measurements. The mixture was prepared by mixing water and glycerin with a ratio of 4.03:6.07 using a magnetic stirrer for at least 30 minutes, resulting a uniform distribution. The density and dynamic viscosity of a water-glycerin mixture with $60.7w\%$ of glycerin is $1.16 \, \mathrm{g/cm}^3$ and $11.64\times 10^{-3}$ Pa s, respectively \cite{cheng_formula_2008}. This yields a negligible gravity-induced velocity of the seeding particles based on a calculation that assumes Stoke's drag \cite{stokes1851} ($v_s=1.14 \,\mathrm{\mu m / s}$). The ability of the particles to follow the fluid flow is evaluated by the relaxation time which can also be calculated based on Stoke's drag force \cite{stokes1851}. For the used fluid and particle, the relaxation time is $\tau _{p}=0.1 \, \mu$s, which is orders of magnitude smaller than the characteristic flow timescale, defined as the tube expansion duration which is on the order of a few hundred milliseconds.
The particles were identified and tracked using the open source, Python-based Trackpy library \cite{allan_soft-mattertrackpy_2025}. The eccentricity of the tubes, although small, introduced geometric asymmetry, producing different cross-sectional areas above and below the inner tube and, consequently, each region should be analysed independently. Therefore, the present study reports results from both areas of the annulus, where the bottom sections is slightly wider than the top section (Fig.~\ref{fig:raw_tubes_image}). However, the effect of eccentricity on the flow properties will be discussed in a future work which is currently under way.

\subsubsection{Pulse wave amplitude}

\begin{figure}[h]
    \centering
    \includegraphics[width=1\linewidth]{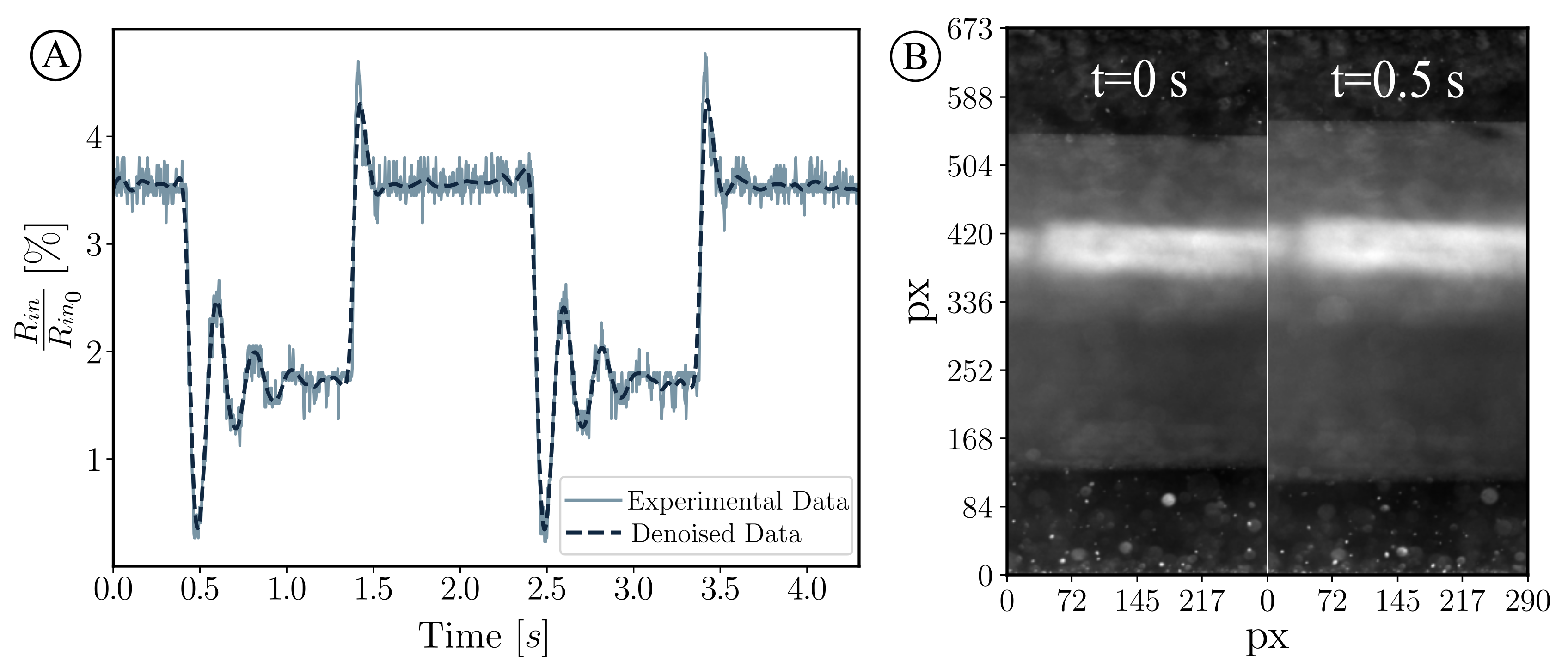}
    \caption{Compliant tube response to pressure pulses. A) Response in time to pressure pulses of $190$ kPa and frequency of $0.5$ Hz. B) Images of the compliant tube captured at the pulse onset (left) and at the pulse midpoint (right) for a $190$ kPa and $1$ Hz pressure pulses, demonstrating the tube dilation.}
    \label{fig:latex-tube-res}
\end{figure}

In addition to investigating the flow in the annular domain, we are also interested in determining the amplitude of the pulse wave and its velocity. These wave properties strongly affect the observed peristaltic flow and its characteristics.
The Pulse Wave Amplitude (PWA) is evaluated by detecting the tube boundaries using image analysis. The tube width in each frame was determined by averaging the distance between the tube boundaries over a short segment of $20$ pixels, where the distance was assumed to remain constant. Figure \ref{fig:latex-tube-res}A demonstrates the tube's response in time to a periodic pulse wave, resulting in a square-like wave with elastic recovery during the contraction. PTV recordings of the tube dilation are presented in Fig.~\ref{fig:latex-tube-res}B. The dilation rate of the compliant tube in response to the pulse wave is significantly affected by its axial tension. Additionally, due to a pressure gradient along the tube, the exact measured location with respect to the valve position could also impact the tube width. Therefore, the PWA should be measured and evaluated individually in every experiment performed.

\subsubsection{Pulse wave velocity and wavelength}


 The expected Pulse Wave velocity (PWV) along the compliant tube can be estimated using the Bramwell-Hill equation for PWV \cite{bramwell_velocity_1922} (re-derived in Appendix \ref{app:pwv-model})
\begin{equation} \label{eq:pwv-intext}
    C=\sqrt{\frac{A_{0}}{\rho}\, \frac{\partial P}{\partial A}}\,\, ,
\end{equation}
where $\rho$ is the fluid density, $A_{0}$ is the nominal cross-sectional area of the tube, and $A$ is the cross-sectional area when internal pressure $P$ is applied. Using the relation between $P$ and $A$ provided by Quarteroni \& Formaggia~\cite{quarteroni2004cardiovascular}, we obtained (full derivation presented in Appendix \ref{app:pwv-model}):
\begin{equation}\label{eq:DP/DA}
    \frac{\partial P}{\partial A}=\frac{\sqrt{\pi}\,\delta\,E}{2A_{0}^{3/2} \, (1-\nu^{2})}\,\, ,
\end{equation}
where $\delta$ is the tube wall-thickness and $E$ and $\nu$ are the Young's modulus and Poisson's ratio of the tube, respectively. The expected PWV in our setup, based on Eq.~\eqref{eq:pwv-intext} and Eq.~\eqref{eq:DP/DA}, is approximately $29 \, \mathrm{m/s}$.

\begin{figure}[h]
    \centering
    \includegraphics[width=1\linewidth]{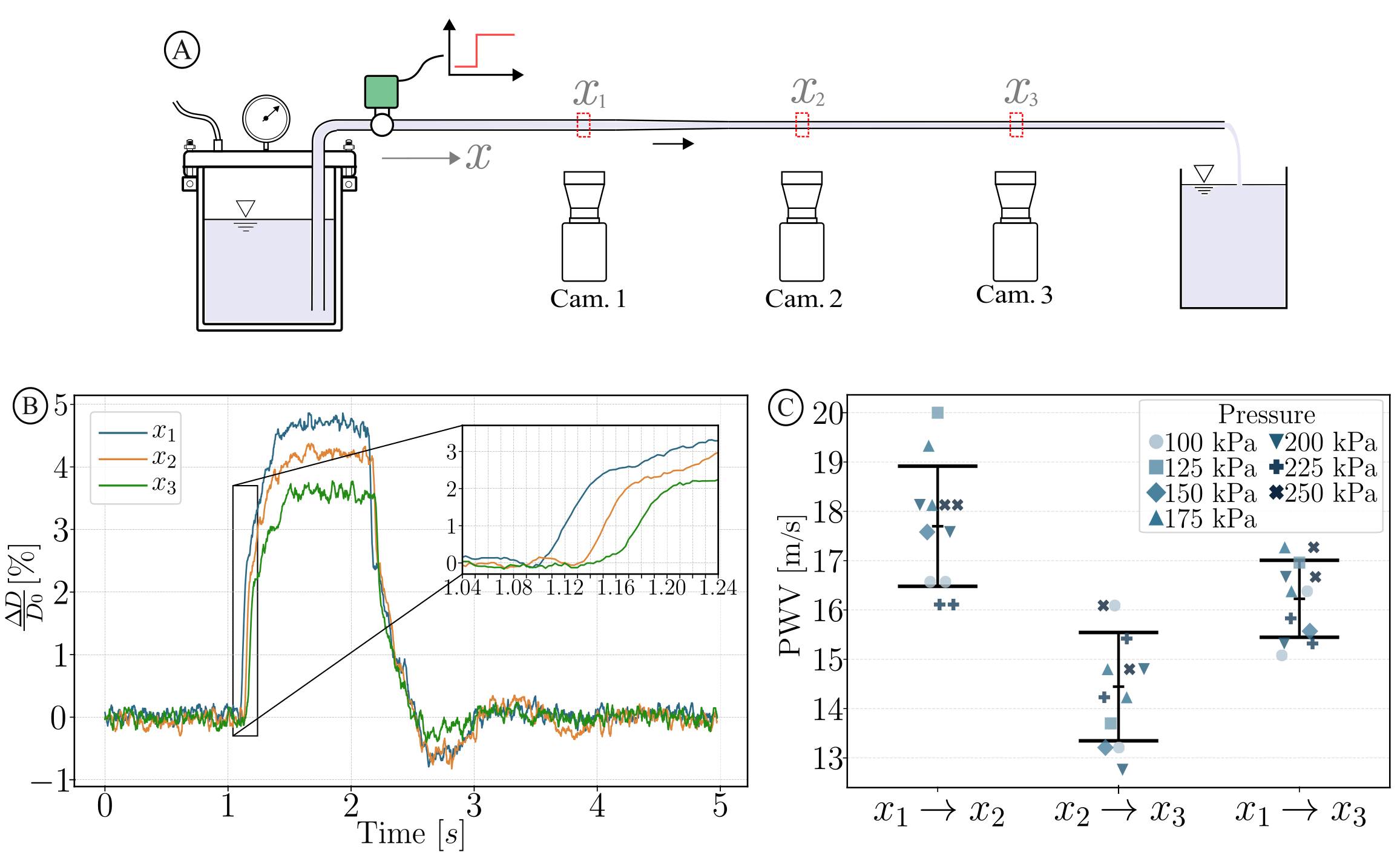}
    \caption{PWV measurements in the complaint tube. A) Schematic sketch of the three-cameras setup used to measure the PWV in the compliant tube. B) Compliant tube response to a $100$ kPa pressure pulse, measured in three locations along the tube. The observed temporal offsets in expansion onset across the different sites support the presence of a propagating wave. C) PWV values measured for different pulse pressure values, showing no dependency on the pressure. Error bars represent the mean $\pm$ standard deviation.}
    \label{fig:pwv_panel}
\end{figure}

To verify that the pulse wave propagates in the compliant tube with a velocity similar to that predicted by Eq.~\eqref{eq:pwv-intext} and Eq.~\eqref{eq:DP/DA}, we conducted a separate experiment. We used a one meter-long compliant latex tube, which was fixed horizontally, and three high-speed cameras (Redwood, IO industries Inc.) that were positioned at three locations along the tube: $x_{1}=0$ mm (reference location), $x_{2}=580$ mm and $x_{3}=950$ mm (Fig.~\ref{fig:pwv_panel}A). The cameras were synchronized such that the tube width could be evaluated at the same time at the three positions. A single pulse was applied by opening the solenoid valve for $1$ s and we recorded the tube's response simultaneously at all three locations. We measured PWVs over the range of $100–250$ kPa, in increments of $25$ kPa, performing two repetitions at each pressure level. As for the PWA, we used image analysis and calculated the PWV based on the tube expansion time in each camera location and the distance between the cameras.

In Fig.~\ref{fig:pwv_panel}B we show the change in the compliant tube diameter at the three measurement locations following a pulse wave generation as a function of time. The diameter increases rapidly once the pulse reaches each measurement location and the PWV can be evaluated based on the time delay of the diameter increase (shown in the inset). Based on these measurements, the PWV for our tube in air (not inside the rigid tube as in the flow measurement experiments), was in the range of $\sim 16-20$ m/s for $x_1 \rightarrow x_2$ and $\sim 12-16$ m/s for $x_2 \rightarrow x_3$. The decrease in PWV was not expected and might be explained by the gradient of water pressure along the tube that was not taken into account in the analytical model. The PWV we measured is lower by 31-59\% as compared to the values estimated using Eq.~\eqref{eq:pwv-intext}, which we associate with uncertainty regarding the complaint tube material properties, as we did not measure them directly, and the fact that we can only accurately measure the external cross section of the tube during the experiment. This PWV corresponds to a wavelength in the range of 5 to 10 meters. Therefore, the peristaltic flow in the presented experimental apparatus is characterized by a wavelength that is approximately two orders of magnitude larger than the characteristic length scale of the flow domain in our setup. This is compatible with the case for CSF flow in cerebral PVSs \cite{kelley_cerebrospinal_nodate, thomas_fluid_2019}.
We also found the PWV not to depend on the pressure level itself as depicted in Fig.~\ref{fig:pwv_panel}C; according to Eq.~\eqref{eq:pwv-intext}, this is a result of $\frac{\partial P}{\partial A}$ being constant within our measurement range. It can also be seen that the maximal change in diameter depends on the measurement location and decreases by approximately 1\% within a distance of one meter. Therefore, as the rigid tube is much shorter than that, axial changes of tube diameter in the flow measurement experiments due to pressure gradient are negligible.

\section{Peristaltic flow measurements}

\subsection{Instantaneous flow}

\begin{figure}[h]
\centering
\includegraphics[width=1\linewidth]{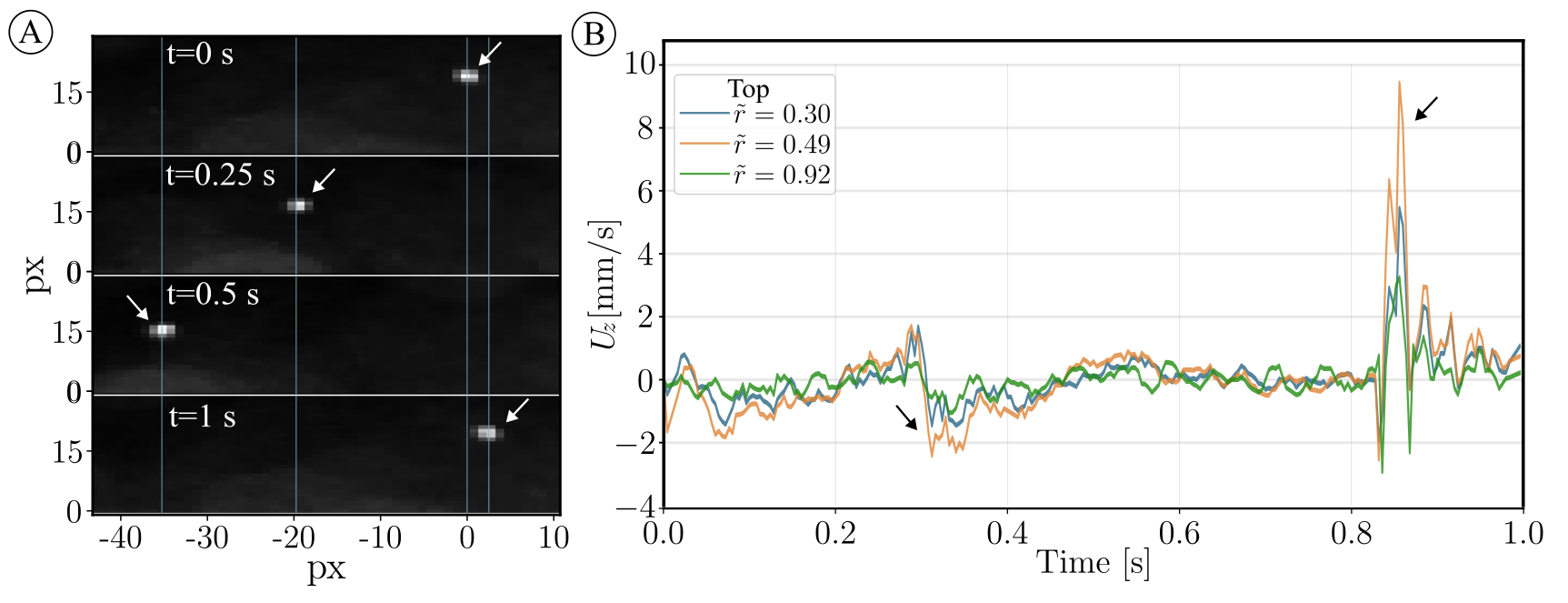}
\caption{Instantaneous flow during full cycle of $50$ kPa pressure pulse at $1$ Hz. A) The motion of a tracer particle with $\tilde{r}=0.47$ at four time points over the course of one cycle. B) The instantaneous velocities of three particles with different $\tilde{r}$ values in the top section of the annulus, exhibiting similar behaviour with magnitudes that vary with $\tilde{r}$.}
\label{fig:instant-velocity}
\end{figure}

The pulse wave travelling along the inner compliant tube was found to induce a complex axial, back and forth flow motion inside the annular gap, yet with a net forward motion in the direction of pulse wave propagation. This motion is demonstrated in Fig.~\ref{fig:instant-velocity}A through the displacements undergone by one tracer particle found approximately at the centre of the annular gap throughout one pulse cycle. During the first half of the cycle the particle moves in the negative direction (opposite to the pulse wave propagation). Then, in the second half of the cycle, the particles displaces in the positive direction (in the direction of pulse wave propagation), eventually passing its initial location. That the particle had passed its initial position shows that there is net fluid motion in the direction of pulse wave propagation throughout the cycle.

The flow in the annular gap was time dependent and the velocity changed with the radial position. Time series of the instantaneous velocity of three particles is shown in Fig.~\ref{fig:instant-velocity}B taken at different radial positions in the top section of the annular gap. The radial position here is normalized by the gap width and is measured relative to the boundary of the inner tube in the neutral pressure conditions ($R_{in_{0}}$),
\begin{equation}
	\tilde{r}=\frac{r-R_{in_{0}}}{R_{out}-R_{in_{0}}}\,\, .
\end{equation}
There are two primary peaks (marked with black arrows), one with a negative velocity at $\sim0.32$ s and one with a positive velocity at $\sim0.86$ s. The peak values are higher for the particles closer to the centre of the annular gap ($\tilde{r}=0.5$). For example, the maximal positive velocity for the particle at $\tilde{r}=0.49$ is $9.41$ mm/s, while for particle at $\tilde{r}=0.92$ it is $3.17$ mm/s. The ratio between the maximal forward velocity to the maximal backward velocity is in the range of 3-4 for all $\tilde{r}$ values we tested.

\subsection{Net flow profile}

To address the net motion of the fluid, without regarding its instantaneous back and forth motion, we captured images of the flow tracers at the same rate at which pulses were generated. This allowed us to track particles at intervals of whole pulse cycles and thus obtain an estimate of the net flow velocity. 
We define the net flow velocity as the translation of a tracer particle over one pulse cycle divided by cycle period ($T$),
\begin{equation}\label{eq:global_velocity}
    \langle \vec U \rangle \equiv \frac{\vec x(t+T) - \vec x(t)}{T} \,\, .
\end{equation}

\begin{figure}[h!]
	\centering
	\includegraphics[width=1\linewidth]{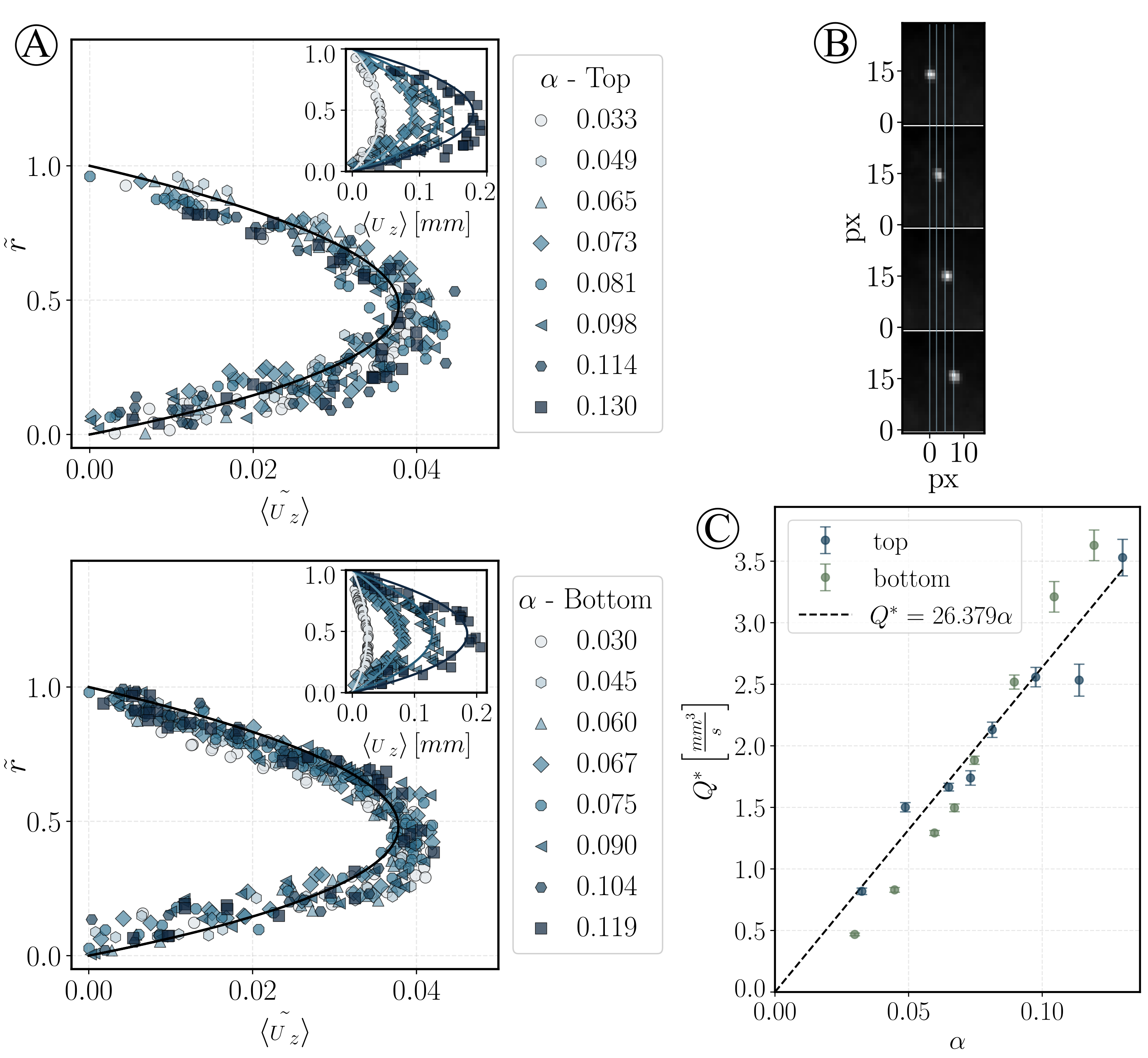}
	\caption{Measured net flow profiles. A) Four representative profiles for different $\alpha$ values at $1$ Hz (inset). The main figure shows the net velocity profiles for the various $\alpha$ values as a function of the normalized radial coordinate. Each data point represents the net velocity of a single particle, normalized by the fitted pressure gradient of the analogue analytical profile and the area of the annular cross section. B) The net forward motion of a particle with $\tilde{r}=0.47$ across four full cycles of $50$ kPa pressure pulses. C) The volumetric flux calculated by fitting the measured velocity profiles to Eq.~\eqref{eq: analytical-flow-profile}, shown as a function of the normalized PWA.}
	\label{fig:flow-profiles}
\end{figure}

Here we will focus on the axial component of the net velocity, $\langle U_z\rangle$. The net axial flow velocity was estimated for multiple particles in our experiment at various normalized radial positions, $\tilde{r}$, which resulted in an estimate of the net velocity profile $\av{U_z(\tilde r)}$.
We repeated this measurement for several pulse amplitudes PWA as shown in the inset of Fig.~\ref{fig:flow-profiles}A, each case specified by the normalized pulse amplitude
\begin{equation}
	\alpha=\frac{R_{in_{max}}- R_{in_{0}}}{R_{out} - R_{in_{0}}}.
\end{equation}
For all cases tested the velocity peaked near the centre of the annular gap and gradually decreased and reached zero at the tube boundaries. The maximum net velocity increased with $\alpha$. Notably, the net velocity was approximately two order of magnitude smaller than the maximal instantaneous velocity. This net axial motion is a result of the repetitive forward motion shown in Fig.~\ref{fig:flow-profiles}B.

To quantify the resulted velocity profiles, we consider a simplified analogy in which the net flow properties are compared to a steady-state laminar flow of a Newtonian fluid, driven by a pressure gradient in-between two fixed concentric tubes. Solving the governing equations for this case (Navier Stokes equations) with no slip boundary conditions, the analytical velocity profile (derived in Appendix \ref{app:annular-flow}) is:
\begin{equation} \label{eq: analytical-flow-profile}
    U^{*}_{z}(r) =\frac{1}{4\mu}\frac{\partial p}{\partial z}^{*}\left[r^2 + \left( \frac{R_{in}^{*^{2}}-R^2_{out}}{ln \frac{R_{out}}{R_{in}^*}} \right) \ln r + \left( \frac{R^2_{out}\ln R^{*}_{in}-R_{in}^{*^{2}}\ln R_{out}}{ln \frac{R_{out}}{R_{in}^*}} \right) \right]\,\, ,
\end{equation}
where $\mu$ is the fluid viscosity, $\frac{\partial p}{\partial z}^{*}$ is the analogous axial pressure gradient, $R_{in}^{*}$ is a representative value of the time-dependent radius of the internal tube and $R_{out}$ is the fixed radius of the external tube.
In the experimental setup presented here, the length of the annular channel is considerably smaller then the peristaltic wavelength and there is no pressure difference between the inlet and the outlet. Consequently, during most of the cycle, there is no pressure gradient along the annular channel, aside from a brief period between the entry and exit of the pulse front and the pulse tail, which is on the order of a few milliseconds. Although this period is short relative to the overall cycle duration, we hypothesize that it is responsible for the observed net flow in the direction of the wave propagation. The axial pressure gradient in the laminar flow analogy, $\frac{\partial p}{\partial z}^{*}$, can be interpreted as the effective pressure gradient that would be required to drive a pressure-driven steady flow of the same discharge.

Using the experimental data of $\av{U_{z}(r)}$ as $U_{z}^{*}(r)$, we found the value of $ \frac{1}{4\mu} \frac{\partial p}{\partial z}^{*}$ for each $\alpha$ value that minimizes the error between the experimental measurements and Eq.~\eqref{eq:global_velocity} using least squares method. For this calculation, we used the nominal radius of the compliant tube, $R_{in_{0}}$, as $R_{in}^{*}$ and assumed a concentric case ($\varepsilon = 0$). The simplified analytical model is plotted for several $\alpha$ values in the inset of Fig.~\ref{fig:flow-profiles}A. The velocity profiles for all $\alpha$ values, each normalized by the associated $\frac{1}{4\mu} \frac{\partial p}{\partial z}^{*}$ and the area of the annular cross section, are shown in the main panels of Fig.~\ref{fig:flow-profiles}A. Even though we used a broad range of $\alpha$ values, the experimental data is seen to collapse well for both top and bottom sections of the annular channel and the normalized data is closely approximated by Eq.~\eqref{eq: analytical-flow-profile} seen as a black line. However, it should be noted that the experimentally obtained velocity is higher than the prediction of the analytical model, particularly near the centre. Using the analytical model fit, we calculated the expected net flux for zero eccentricity (Fig.~\ref{fig:flow-profiles}C), with error bars derived from the uncertainties obtained via the least squares method. The flux increased with $\alpha$, and the data were fitted with a linear trend line. The resulting level of agreement with the trend line suggests that the uncertainty in the measurement does not arise solely from the analytical model approximation, but likely also reflects other factors, such as small degrees of three-dimensional motion or eccentricity.
Although both sections of the annular channel followed similar overall trends, with a positive net motion in the direction of the pulse wave propagation, some differences were observed. A slightly higher net velocity was obtained in the top section with a greater scatter relatively to the bottom section. These differences can result from challenges in centring the tubes, different lighting conditions, and other factors that are difficult to predict.

\section{Conclusions}

To conclude, this work presents a novel experimental system designed to emulate peristaltic flow in annular geometries due to periodic propagating pulse waves in the internal tube---a setup relevant for understanding the mechanisms driving the flow of CSF through cerebral PVS~\cite{kelley_cerebrospinal_nodate, thomas_fluid_2019}. Importantly, the setup is fully optically accessible, and thus it allows detailed flow and pulse wave measurements inside the annular gap. Using the new experimental setup, this work demonstrates that pulse waves with a wavelength significantly larger than the tube length can drive a net flow inside the annular gap in the direction of the pulse wave propagation. This observation is consistent with the findings of Coloma et al.~\cite{coloma_boundary_2019}, who reported the same phenomenon in a rectangular channel. We have also demonstrated that the instantaneous flow behaviour is much more complex and vigorous than the net flow transport. Furthermore, through PTV measurements we showed that the net axial velocity profile has a self-similar shape (the profiles collapse across a wide range of PWA we measured) that is reminiscent of the flow profile of a steady laminar flow in annular gaps obtained by solving the Navier Stokes equations for these conditions. These results are in agreement with those of Mestre et al.~\cite{mestre_flow_2018} who found that the time-averaged net flow of CSF in cerebral pial PVS of mice is in the same direction as the blood flow and the velocity profile is nearly parabolic. In future studies, whose experiments are currently under way, we will utilize this new apparatus to explore the features of this flow in more depth. We anticipate that our experimental approach will provide instrumental in validating hypotheses and theories related to CSF transport in the glymphatic system.

Our choice of utilizing particle tracking for this flow measurements has advantages and disadvantages. A significant advantage is the excellent spatial and velocity measurement resolution, as our method can accurately reconstruct tracer particles' trajectories with resolutions down to 1\% of the gap width in the radial and axial coordinates. The disadvantage, on the other hand, is that the out-of-plane (azimuthal) component of the motion, which is expected for asymmetric annular configurations ~\cite{carr_peristaltic_2021}, is difficult to capture and we have not attempted to measure it as of now. A possible way of achieving such three dimensional measurements would be to utilize de-focusing techniques (e.g. Refs.~\cite{Fuchs2016, Afik2017}), for which we will need to further adapt our imaging setup.

The physical scale chosen for our setup presented a compromise between several considerations. On the one hand, a high degree of control over the apparatus geometry is needed to be able to consistently produce the pulses, for which larger designs are more suitable. Larger designs also allow better manipulation of eccentricity and miss-alignment of the annular gap.  Furthermore, utilizing larger designs makes flow measurements of complex velocity fields somewhat more accessible as compared to microscopic flow setups which are limited to narrow two-dimensional planes and limited in spatial extent. On the other hand, peristaltic flows in most physiological contexts are laminar, occurring at very low Reynolds numbers, which can be challenging to achieve in large setups. To compensate over these competing aims, we utilized a millimetre-scale setup in addition to a high viscosity fluid. The level of details achieved in our velocity measurements demonstrate the advantages mentioned above, while the possible finite Reynolds number effects for the high pulse wave amplitudes we used demonstrates the limitations of our system. Performing experiments with higher viscosity fluids will tend to lessen this limitation.

Overall, our work introduces a new approach to study the mechanisms underlying physiological fluid transport as in CSF flow through cerebral PVSs. While our current model captures key dynamics, several factors, such as eccentricity and tube size ratio, remained to be explored and experiments are currently under way to address their effects too. We hope that this system will enhance understanding of transport processes relevant to physiological issues, with potential of improving drug delivery and therapeutic processes.

\subsubsection*{Acknowledgments}

This work was partially funded by ISF grant number 1244/24 and 2586/24. We thank Dr. Nurit Omer for fruitful discussions on CSF flow and neurological-related conditions. We also thank Dr. Manikandan Raghunathan and Naama Greidi for assistance with some of the experiments.

\bibliographystyle{vancouver}
\bibliography{References/General,References/PDMS_and_Porous_Media,References/Pulsatile_Flow,References/Peristaltic_Flow,References/CSF_Flow_Mechanism}

\appendix
\numberwithin{equation}{section}
\counterwithin{figure}{section}
\section*{Appendix}

\section{Pulse Wave Velocity Model} \label{app:pwv-model}

\begin{figure}[h]
\centering
\includegraphics[width=0.5\linewidth]{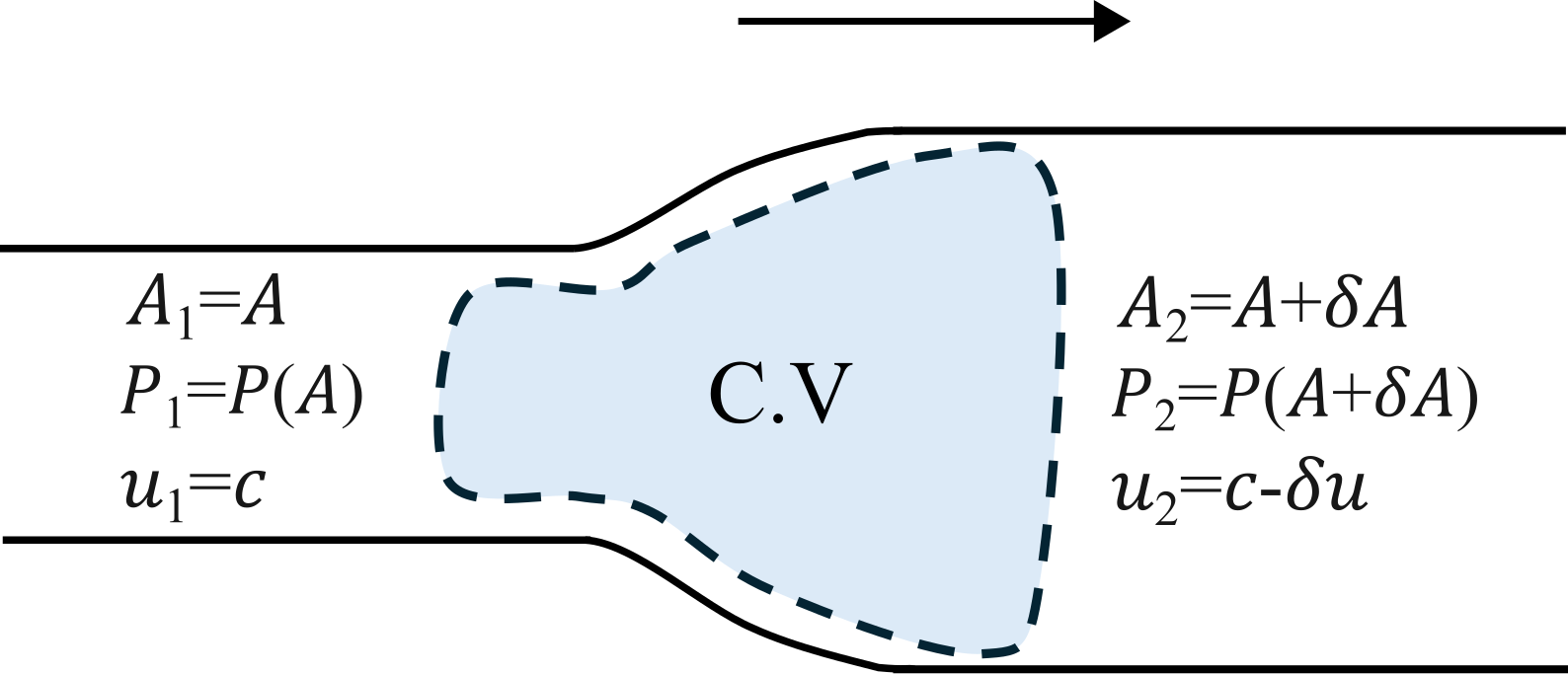}
\caption{Schematic of pulse wave propagation in a compliant tube. A pulse wave travels to the right with wave speed $c$ in a tube of nominal cross-sectional area $A$. Upstream of the wave, the flow has area $A_1 = A$, pressure $P_1 = P(A)$, and axial velocity $u_1 = c$. Within the control volume (C.V), the pulse induces a local dilation of the tube, increasing the cross-sectional area to $A_2 = A + \delta A$. This dilation is accompanied by an increase in pressure to $P_2 = P(A + \delta A)$ and a corresponding decrease in flow velocity to $u_2 = c - \delta u$. Here, $u$ and $P$ denote the fluid velocity and pressure, respectively, and $\delta A$ and $\delta u$ represent small perturbations associated with the pulse.}
\label{fig:pwv-sketch}
\end{figure}

We can derive the pressure-dependent term of PWV \cite{bramwell_velocity_1922} using a controlled-volume approach as presented in Fig.~\ref{fig:pwv-sketch}.
Assuming uniform velocities at the inlet and outlet of the control volume ($u_1$ and $u_2$ respectively), mass conservation leads to:
\begin{equation}
    A_{1}u_{1}=A_{2}u_{2}.
\end{equation}
where $A_1$ and $A_2$ are the cross sectional areas of inlet and outlet of the control volume.

Substituting the terms as stated in Fig.~\ref{fig:pwv-sketch} gives:
\begin{equation}
    c=\delta u \frac{A+\delta A}{\delta A},
\end{equation}

and for small changes, where $A \gg \delta A$, we can write:
\begin{equation}
    c=\delta u \frac{A}{\delta A}.
\end{equation}

From momentum conservation we have:
\begin{subequations}
\begin{gather}
    \int \rho \vec{u}( \vec{u} \vec{\hat{n}})dA-\sum F=0 \\
    \rho (u^{2}_{1}A_{1}-u^{2}_{2}A_{2})-(P_{2}A_{2}-P_{1}A_{1})=0 \\
    \rho u_{1}A_{1}(u_{1}-u_{2})-(P_{2}A_{2}-P_{1}A_{1})=0.
\end{gather}
\end{subequations}

Using the terms from Fig.~\ref{fig:pwv-sketch} we get:
\begin{equation}\label{eq: wave-vel-1}
    \rho Ac\delta u-[P(A+ \delta A) (A+ \delta A)-P(A) A]=0.
\end{equation}

Rearranging Eq.~\eqref{eq: wave-vel-1} yields the partial derivative of $PA$ with respect to $A$:
\begin{equation}
    \rho c^{2}=\frac{P(A+ \delta A) (A+ \delta A)-P(A) A}{\delta A}=\frac{\partial (PA)}{\partial A}.
\end{equation}

Using the chain rule, we can write the term for the pulse wave velocity:
\begin{equation}
    c=\sqrt{\frac{P}{\rho}+\frac{A}{\rho} \frac{\partial P}{\partial A}},
\end{equation}

and for small cross-section changes we have:
\begin{equation}
    c=\sqrt{\frac{A}{\rho} \frac{\partial P}{\partial A}}.
\end{equation}

By simplifying the linear elastic law, Quarteroni \& Formaggia \cite{quarteroni2004cardiovascular} derived a relation for $P$ and $A$:

\begin{equation}
P=\frac{\sqrt{\pi}tE}{1-\nu^2}\cdot \frac{\sqrt{A_{p}}-\sqrt{A}}{A},
\end{equation}

 where $t$ is the tube thickness, $E$ and $\nu$ are the Young's modulus and Poisson's ratio of the tube, respectively, and $A_p$ is the expanded cross-section area of the tube. Calculating the partial derivative with respect to $A$ gives:

 \begin{equation}
    \frac{\partial P}{\partial A}=\frac{\sqrt{\pi}tE}{2A\sqrt{A_{p}}(1-\nu^{2})} .
\end{equation}

For $A_{p}-A\ll 1$ we get the approximation:
 \begin{equation}
    \frac{\partial P}{\partial A}=\frac{\sqrt{\pi}tE}{2A^{3/2}(1-\nu^{2})}.
\end{equation}

\section{Annular Flow Model}\label{app:annular-flow}

The governing equation for a steady and laminar axial flow of an incompressible Newtonian fluid in cylindrical coordinates is
\begin{equation}
    \mu \frac{1}{r} \frac{\partial}{\partial r} \left(
    r\frac{\partial u_z}{\partial r}
    \right) = \frac{\partial p}{\partial z}
\end{equation}
The exact solution for this equation is \cite{kundu_fluid_2016}:
\begin{equation}
    u_{z}(r)=\frac{r^2}{4\mu}\frac{\partial p}{\partial z}+A\ln r+B,
\end{equation}

where $\mu$ is the dynamic viscosity of the fluid and $p$ is the pressure inside the tube.
In the case of flow between two concentric tube, where $r \in [R_{in}, R_{out}]$, the no-slip boundary conditions are
\begin{equation}
    \left\{\begin{array}{rc}u_{z}(r=R_{in})=0 \\ u_{z}(r=R_{out})=0 \end{array}\right. .
\end{equation}

Thus
\begin{subequations}
\begin{align}
    A&=\frac{1}{4\mu}\frac{\partial p}{\partial z}\left( \frac{R^2_{in}-R^2_{out}}{ln \frac{R_{out}}{R_{in}}} \right) \\
    B&=\frac{1}{4\mu}\frac{\partial p}{\partial z}\left( \frac{R^2_{out}\ln R_{in}-R^2_{in}\ln R_{out}}{ln \frac{R_{out}}{R_{in}}} \right).
\end{align}
\end{subequations}

Peristaltic flow between two concentric tubes is time depended, as $\frac{\partial p}{\partial z}=\frac{\partial p}{\partial z}(t)$ and $R_{in}=R_{in}(t)$ (As in the experiment, we neglect axial changes in $z$). However, in this work we consider an effective pressure gradient, $\frac{\partial p}{\partial z}^{*}$, that represents the driving mechanism of the peristaltic pumping analogous to the empirically measured flow:
\begin{equation}
    u^{*}_{z}(r)=\frac{1}{4\mu} \frac{\partial p}{\partial z}^{*} \left[r^2 + \left( \frac{R_{in}^{*^2}-R^2_{out}}{ln \frac{R_{out}}{R_{in}^{*}}} \right) \ln r + \left( \frac{R^2_{out}\ln R_{in}^*-R_{in}^{*^2}\ln R_{out}}{ln \frac{R_{out}}{R_{in}^*}} \right) \right].
\end{equation}

\end{document}